\documentclass[aoas,nameyear,MSNbibl,dvips]{arximspdf}

\doi{10.1214/10-AOAS398G}
\referstodoi{10.1214/10-AOAS398}
\volume{5}
\issue{1}
\pubyear{2011}
\firstpage{61}
\lastpage{64}

\makeatletter

\def\bptnote#1{}

\makeatother

\begin{document}
\begin{frontmatter}

\title{Discussion of: A statistical analysis of multiple
temperature proxies: Are reconstructions of surface temperatures over the last\\ 1000
years reliable?}
\runtitle{Discussion}
\pdftitle{Discussion on A statistical analysis of multiple temperature proxies:
Are reconstructions of surface temperatures over the last 1000 years reliable?
 by B. B. McShane and A. J. Wyner}

\begin{aug}
\author[A]{\fnms{Murali} \snm{Haran}\corref{}\ead[label=e1]{mharan@stat.psu.edu}}
and
\author[B]{\fnms{Nathan M.} \snm{Urban}\ead[label=e2]{nurban@princeton.edu}}

\runauthor{M. Haran and N. M. Urban}

\affiliation{Pennsylvania State University and Princeton University}

\address[A]{Department of Statistics\\
Pennsylvania State University\\
University Park, Pennsylvania 16802\\
USA\\
\printead{e1}}

\address[B]{Woodrow Wilson School of Public \\
and International Affairs\\
Princeton University\\
Princeton, New Jersey 08544\\
USA\\
\printead{e2}}
\end{aug}

\received{\smonth{9} \syear{2010}}



\end{frontmatter}

We thank the authors for a thought-provoking paper (henceforth MW).
Their work may be divided into two parts: \textit{reconstruction}, where
the authors develop a Bayesian model for reconstructing historic
temperatures based on proxies, along with associated measures of
uncertainty; and \textit{validation}, where they study how accurately
their model corresponds to data by using cross-validation techniques
or comparing proxies to simulated time series that are unrelated to
temperature.  We discuss both aspects of the paper although we focus
mostly on reconstruction. While our comments may seem critical of MW,
our views apply more generally to much of the existing work in this
area.

We begin with a discussion of the reconstruction in MW. Given the
advances in modeling for large, rich, complicated space--time processes
and the availability of temperature proxies in the form of space--time
data sets, we believe statistical approaches to paleoclimate
reconstruction should make full use of such spatial data instead of
using spatially aggregated forms of the data (as in MW).  Such spatial
aggregation may have the effects of removing interesting signals and
of making it more difficult to define a credible error structure since
proxy data are less directly related to global temperature than local
temperature. This is an issue not only with MW, but also the
reconstruction work of many others. In addition, recent advances in
computationally efficient approaches for fitting hierarchical
spatiotemporal models open up the possibility of developing more
realistic models that account for various sources of error while
incorporating specialized scientific knowledge into the models as
appropriate [cf.  \citet{bancarlgelf2004}; \citet{GelfDiggFuenGutt2010} and the references
therein]. We believe that
such models are likely to provide more reliable estimates along with
associated uncertainty estimates, both of which are important for
drawing sound scientific conclusions.

We outline some ways in which we believe the model in MW can be
improved upon.

 (i) The authors approach this as a regression problem
where they treat the proxies as the predictor and the temperature
observations as response, and then use the proxies to extrapolate the
temperature backwards. We believe\vadjust{\goodbreak} it is more appropriate to view
temperature as predicting proxies rather than the other way around.
Recasting the problem in this ``calibration'' framework allows for
more realistic models for measurement error and dependence. As is well
known, ignoring measurement error in regressions can lead to erroneous
conclusions [cf. \citet{fuller1980properties}; \citet{carroll1995measurement}].

 (ii) The process by which MW selects proxies is problematic. We
wonder why MW choose only those proxies that go all the way back given
the availability of approaches for dealing with missing information.

 (iii) The
proxies are all very different in terms of scale, how they were
collected and possibly aggregated, the kind of measurement error
involved, and other characteristics such as spatial dependence.  Also,
the data associated with the proxies may be very different; for
instance, some are discrete, some are continuous, and they may be at
different frequencies.  Critically, proxies are vastly different in
terms of what they tell us about temperature; for example boreholes
provide highly ``smoothed'' temperature reconstructions, while tree
rings or lake varves can have annual resolution. It therefore seems
inappropriate to merge all proxies together into a single
regression model without accounting for their individual properties.

(iv) Principal Components Analysis (PCA) is a reasonable approach to
reducing the dimensions of predictors in a regression problem, but we
are concerned that PCA, like the LASSO, treats the
paleo-reconstruction problem like a ``data mining'' problem, that is,
a problem where nothing is known about underlying relationships among
the predictors and the temperature field. For instance, negative
regression coefficients may not be tenable in several cases.

(v) It
may be possible to construct more realistic proxy-temperature
relationships using process models [cf. \citet{Guiot2009}], although
this may be more feasible for stronger climate signals, for example,
deglaciations, than those present in the late Holocene.

We agree with the authors that if proxies show no recent changes then
they may be inappropriate for extrapolating backwards.  However, any
discussion of Holocene paleoclimate reconstructions should be kept in
perspective.  It is tempting to use paleodata to make inferences about
future climate.  There have been attempts to use Holocene paleodata to
constrain the climate sensitivity [cf. \citet{Hegerl2006}; \citet{schneider2007climate} for discussions of
related methodology], though they do
not account for temporal or spatial dependence, and share the
limitations of the paleo-reconstructions upon which they are based.
But independent of the accuracy of a paleotemperature reconstruction,
there are limits to what past climates can tell us about possible
future climates.  Over the next few centuries the climate system will
likely be strongly forced by continued greenhouse gas emissions.  By
contrast, the Holocene climate was relatively weakly forced, and not
primarily by greenhouse gases.  Given these differences, it is unclear
to what extent further refinement of millennial temperature
reconstructions can contribute to questions about the future
climate.\vadjust{\goodbreak}
 However, this does not detract from their potential usefulness in
answering questions about natural climate variability, such as
spatiotemporal patterns and attribution of past climate change.

Regarding the validation approach in MW: while we appreciate the principle
behind comparing proxy-based reconstructions with constructions based
on randomly generated proxies, it is perhaps not entirely surprising
that models with dependent errors are good interpolators over short
time periods. The actual proxies themselves may not be as good for
short time periods, especially in the case of low-frequency proxies
like a borehole, or an ecological proxy like a tree ring which might
be confounded by subdecadal non-temperature variability [this
is related to issue (iii) above]. One might also believe that a proxy
would perform better at extrapolation over longer time periods.

In summary, we do not argue for or against the conclusions of this
paper as much as we argue that much of the statistical work done in
this paper and other related papers do not take full advantage of
existing data, scientific knowledge and the latest in statistical
methods, particularly hierarchical space--time modeling [see \citet{tingleyetal2010} for
a discussion of possible strategies to pursue].
Having said that, the researchers in this field deserve much credit
for their pioneering work on temperature reconstructions which has
laid the foundations for an important and interesting field of
research.
We are delighted that more statisticians are becoming involved in the
statistical aspects of climate science and we commend the authors for
taking on this challenging problem in a methodical fashion. We
particularly like their method of carefully working through both
reconstruction and validation; this two-pronged approach provides a
nice template for future work.

\section*{Acknowledgment}

We thank Don Richards for helpful discussions.


\printaddresses

\end{document}